\documentclass{ws-procs975x65}            

\newcommand{\be}{\begin{equation}}
\newcommand{\ee}{\end{equation}}
\newcommand{\bea}{\begin{eqnarray}}
\newcommand{\eea}{\end{eqnarray}}
\newcommand{\beaa}{\begin{eqnarray*}}
\newcommand{\eeaa}{\end{eqnarray*}}

\newcommand{\Eqn}[1]{&\hspace{-0.2em}#1\hspace{-0.2em}&}

\def\Vec#1{\mbox{\boldmath $#1$}}

\begin{document}
\title{Large-scale magnetic fields, non-Gaussianity, and gravitational waves from inflation}

\author{Kazuharu Bamba$^{1, 2, 3, 4}$}

\address{
$^1$Leading Graduate School Promotion Center, 
Ochanomizu University, 2-1-1 Ohtsuka, Bunkyo-ku, Tokyo 112-8610, Japan\\
$^2$Department of Physics, Graduate School of Humanities and Sciences, Ochanomizu University, Tokyo 112-8610, Japan\\ 
$^3$Kobayashi-Maskawa Institute for the Origin of Particles and the
Universe, Nagoya University, Nagoya 464-8602, Japan\\
$^4$Division of Human Support System, Faculty of Symbiotic Systems Science, Fukushima University, Fukushima 960-1296, Japan\footnote{The author's current affiliation}\\
E-mail: bamba@sss.fukushima-u.ac.jp
}

\begin{abstract}
We explore the generation of large-scale magnetic fields 
in the so-called moduli inflation.  
The hypercharge electromagnetic fields couple to not only 
a scalar field but also a pseudoscalar one, 
so that the conformal invariance of the hypercharge electromagnetic fields can be broken. 
We explicitly analyze the strength of the magnetic fields on the Hubble horizon scale at the present time, the local non-Gaussianity of 
the curvature perturbations originating from the massive gauge fields, and the tensor-to-scalar ratio of the density perturbations. 
As a consequence, we find that the local non-Gaussianity and the tensor-to-scalar ratio are compatible with the recent Planck results. 
\end{abstract}


\bodymatter

\section{Introduction}

Magnetic fields with the current strength $\sim 10^{-6}$G 
on $1$--$10$kpc scale and 
those with $10^{-7}$--$10^{-6}$G on $10$~kpc--$1$Mpc scale 
have been detected in galaxies and clusters of galaxies, 
respectively. 
The generation mechanism of the large-scale 
magnetic fields in clusters of galaxies 
have not been understood well.~\cite{Kronberg:1993vk,Grasso:2000wj,Widrow:2002ud,Giovannini:2003yn,Kandus:2010nw,Yamazaki:2012pg,Durrer:2013pga}

Electromagnetic quantum fluctuations generated during 
inflation are considered to be the most natural origin of 
such large-scale magnetic fields. 
This is because inflation extends 
the coherent scale of the magnetic fields to be larger 
than the Hubble horizon.~\cite{Turner:1987bw} 
The homogeneous and isotropic 
Friedmann-Lema\^{i}tre-Robertson-Walker (FLRW) universe 
is conformally flat. 
In addition, the electromagnetic fields are conformally invariant. 
Accordingly, the conformal invariance of the electromagnetic fields has to 
be broken at the inflationary stage. 
As a result, the quantum fluctuations of the electromagnetic fields can be 
generated.~\cite{Parker:1968mv,Turner:1987bw} 
As representative mechanisms to break the conformal invariance of the electromagnetic fields, there have been proposed 
the non-minimal coupling of the electromagnetic fields to 
the scalar curvature,~\cite{Drummond:1979pp,Turner:1987bw,Bamba:2006ga,Bamba:2008ja,Bamba:2008xa} 
its coupling to the scalar fields,~\cite{Ratra:1991bn,Garretson:1992vt,Bamba:2003av,Bamba:2004cu,Martin:2007ue} 
and the trace anomaly.~\cite{Dolgov:1993vg} 

In this paper, we review our main results in Ref.~\citenum{Bamba:2014vda}. 
We explore a toy model of moduli inflation inspired by racetrack 
inflation~\cite{BlancoPillado:2004ns} 
in the framework of the Type IIB string theory. 
The new point of our model is to take into account 
the coupling of the hypercharge electromagnetic fields 
to a scalar field as well as to a pseudoscalar one, 
which plays a role of the inflaton field. 
Only the latter coupling has been studied in the past 
works.~\cite{Barnaby:2010vf,Barnaby:2011vw,Meerburg:2012id,Linde:2012bt} 
We explicitly estimate the values of current magnetic fields on the 
Hubble horizon scale, 
local non-Gaussianity of the curvature perturbations, 
and tensor-to-scalar ratio of density perturbations. 
This is the most significant consequence of the present work. 
We use the units $k_{\mathrm{B}} = c = \hbar = 1$ and describe the 
Newton's constant by 
$G=1/M_\mathrm{P}^2$, where $M_\mathrm{P} =2.43 \times 10^{18}$ GeV 
is the reduced Planck mass. 
In terms of electromagnetism, we adopt Heaviside-Lorentz units. 

The paper is organized as follows. 
In Sec.\ 2, the model Lagrangian is explained, and the field equations 
are derived. 
In Sec.\ 3, we investigate the present magnetic field strength on the Hubble horizon scale. 
Furthermore, we study the non-Gaussianity of the curvature perturbations 
in Sec.\ 4, and we explore the tensor-to-scalar ratio of the density 
perturbations in Sec.\ 5. 
Finally, summary is given in Sec.\ 6.

\section{Model Lagrangian} 

The Lagrangian is represented as 
\begin{eqnarray}
{\mathcal L} \Eqn{=}  \frac{M_\mathrm{P}^2}{2} R 
-\frac{1}{4} X F_{\mu\nu}F^{\mu\nu}
-\frac{1}{4} g_{\mathrm{ps}} \frac{Y}{M} F_{\mu\nu}\tilde{F}^{\mu\nu} 
 \nonumber \\
&& {}-\frac{1}{2}g^{\mu\nu}{\partial}_{\mu}{\Phi}{\partial}_{\nu}{\Phi} 
- U(\Phi) 
-\frac{1}{2}g^{\mu\nu}{\partial}_{\mu}{Y}{\partial}_{\nu}{Y} 
- V(Y) \,.
\label{eq:2.1} 
\end{eqnarray} 
Here, $R$ is the scalar curvature, 
$\Phi$ is the canonical scalar field, 
$X \equiv \exp \left(-\lambda \Phi/M_P \right)$ 
with a constant $\lambda$, and 
$V(Y) \approx \bar{V} - \left(1/2\right) m^2 Y^2$, 
where $\bar{V}$ is a constant and $m$ is the mass of $Y$. 
Moreover, $U(\Phi)$ and $V(Y)$ are the potentials of $\Phi$ and $Y$, respectively, $g_{\mathrm{ps}}$ is a dimensionless coupling constant, and 
$M$ is a constant with the mass dimension, which corresponds to the decay constant of $Y$. 
In addition, the field strength of the $\mathrm{U}(1)_Y$ 
hypercharge gauge field $F_{\mu}$ is given by 
$F_{\mu\nu} = \nabla_{\mu} F_{\nu} - \nabla_{\nu} F_{\mu}$ 
with $\nabla_{\mu}$ the covariant derivative, and the dual field strength of $F_{\mu}$ is $\tilde{F}^{\mu\nu}$. 
The pseudoscalar field $Y$ corresponds to the inflaton field. 

We assume the flat FLRW universe with the metric 
${ds}^2 = -{dt}^2 + a^2(t)d{\Vec{x}}^2$, where 
$a$ is the scale factor. 
In this background, 
the field equations of $\Phi$ (i.e., $X$) and $Y$ read 
$\ddot{\Phi} + 3H\dot{\Phi} + dU(\Phi)/d\Phi = 0$ and  
$\ddot{Y} + 3H\dot{Y} + dV(Y)/dY = 0$, respectively, 
where $H \equiv \dot{a}/a$ is the Hubble parameter and 
the dot means the derivative with respect to the cosmic time $t$. 
With the Coulomb gauge 
$F_0(t,\Vec{x}) = 0$ and ${\partial}_j F^j (t,\Vec{x}) =0$, 
we also obtain the field equation of $F_{\mu}$.

\section{Magnetic field strength at the present time} 

For the slow-roll inflation driven by the potential of $Y$, 
the scale factor $a(t)$ can be described by 
$a(t) = a_k \exp \left[ H_{\mathrm{inf}} \left(t-t_k\right) \right]$, 
where $a_k=a(t_k)$, $t_k$ is the time when a comoving wavelength 
$2\pi/k$ of the $\mathrm{U}(1)_Y$ gauge field first crosses the horizon 
during inflation ($k/(a_k H_{\mathrm{inf}}) = 1$), and 
$H_\mathrm{inf}$ is the Hubble parameter during inflation. 
A solution of equation of motion for $Y$ is derived as~\cite{Garretson:1992vt} 
$Y (t) = Y_k \exp \left[ \left(3/2\right) \mathcal{D} H_{\mathrm{inf}} \left( t-t_k \right) \right]$, 
where $\mathcal{D} \equiv -1 \pm 
\sqrt{1 + \left[ 2m/\left(3H_{\mathrm{inf}}\right)\right]}$ 
and $Y_k \equiv Y(t = t_k)$.

We execute the cannonical quantization of the $\mathrm{U}(1)_Y$ gauge field $F_{\mu}(t,\Vec{x})$. 
We express the comoving wave number by $\Vec{k}$ ($k = |\Vec{k}|$). 
We set the $x^3$ axis to lie along the direction of the spatial momentum 
\Vec{k} and express the transverse directions as 
$x^{1}$ and $x^{2}$. 
By using 
the circular polarizations
$F_{\pm}(k,t) \equiv F_1(k,t) \pm i F_2(k,t)$ with 
the Fourier modes 
$F_1(k,t)$ and $F_2(k,t)$ of the $\mathrm{U}(1)_Y$ gauge field, 
we have 
\begin{equation}
\ddot{F}_{\pm}(k,t) 
+ \left( H_{\mathrm{inf}} + \frac{\dot{X}}{X} 
\right) \dot{F}_{\pm}(k,t) 
+ \left[ 1 \pm \frac{g_{\mathrm{ps}}}{M} \frac{\dot{Y}}{X} 
\left( \frac{k}{a} \right)^{-1} 
\right] \left( \frac{k}{a} \right)^2 F_{\pm}(k,t) = 0\,. 
\label{eq:3.12}
\end{equation}
We numerically calculate the solution of this equation 
at the inflationary stage 
by defining the ratio 
$C_{+}(k,t) \equiv F_{+}(k,t)/F_{+}(k,t_k)$, 
where $t_k$ is taken as the initial time of the numerical calculation. 
Here, in the short-wavelength limit of $k \rightarrow \infty$, 
we set $F_{+}(k,t) = 1/\sqrt{2kX(t)}$. 
Namely, we take the so-called Bunch-Davies vacuum so that 
in this limit the vacuum can be the one in the Minkowski space-time. 
As a result, without sensitive dependence of the model parameters, 
$C_{+}(k, t)$ becomes a constant after $\sim 10$ Hubble expansion time. 
after the horizon crossing during inflation. 

The proper hypermagnetic field is expressed as~\cite{Ratra:1991bn} 
${B_Y}_i^{\mathrm{proper}}(t,\Vec{x}) = \left(1/a^2\right) {B_Y}_i(t,\Vec{x}) = \left(1/a^2\right) {\epsilon}_{ijk}{\partial}_j F_k(t,\Vec{x})$, 
where ${B_Y}_i(t,\Vec{x})$ is the comoving hypermagnetic field, and 
${\epsilon}_{ijk}$ is the totally antisymmetric tensor 
(${\epsilon}_{123}=1$).
We suppose that after inflation, the instantaneous reheating happens $t=t_\mathrm{R}$ (much before the electroweak phase transition 
(EWPT), where $T_\mathrm{EW} \sim 100$GeV). 
Carged particles are produced at the reheating stage, 
so that the cosmic conductivity of ${\sigma}_\mathrm{c}$ 
can be much larger than $H$. 
Thus, after reheating, $B_Y$ behaves as $B_Y \propto a^{-2}$. 
On the other hand, 
the hyperelectric fields accelerats the charged particles and 
eventually vanish. 
The energy density of the magnetic fields $\rho_B (L,t)$ 
in the position space reads~\cite{Bamba:2006km}
\begin{equation}    
\rho_B (L,t) \simeq
\frac{1}{8\pi^2} 
\frac{1}{X(t_k)} \frac{1}{\sqrt{2 \xi_k}} 
\exp \left[ 2\left(\pi \xi_k -2\sqrt{2 \xi_k}\right) \right] 
\left( \frac{k}{a} \right)^4 
|C_+(k, t_\mathrm{R})|^2\,, 
\label{eq:3.35}
\end{equation}
with 
$\xi_k = \xi (t = t_k)$, where 
$\xi \equiv g_{\mathrm{ps}} \dot{Y}/ 
\left(2 M X H_{\mathrm{inf}} \right)$. 
Here, we have imposed $X(t_{\rm R})=1$ 
in order or the Maxwell theory is recovered after 
the instantaneous reheating ($t \geq T_\mathrm{R}$). 
We have also neglected the difference between 
the coefficient of the $\mathrm{U}(1)_Y$ magnetic field 
and that of the $\mathrm{U}(1)_{\rm em}$ one, which is $\mathcal{O}(1)$. 

\begin{table}[t]
\tbl{Magnetic field strength on the Hubble horizon scale at the present time.}
{\begin{tabular}{@{}cccccc@{}}
\toprule
& $B(H_0^{-1},t_0) \hspace{1mm} [\mathrm{G}]$  
& $B(1\mathrm{Mpc},t_0) \hspace{1mm} [\mathrm{G}]$  
& $H_{\mathrm{inf}} \hspace{1mm} [\mathrm{GeV}]$ 
& $m \hspace{1mm} [\mathrm{GeV}]$  
& $Y_k/M_\mathrm{P}$
\\[0mm]\colrule
(a)  
& $7.15 \times 10^{-64}$
& $1.42 \times 10^{-56}$
& $1.0 \times 10^{11}$ 
& $2.44 \times 10^{10}$ 
& $7.70 \times 10^{-2}$ 
\\[0mm]
(b)  
& $7.15 \times 10^{-64}$
& $1.42 \times 10^{-56}$
& $1.0 \times 10^{10}$ 
& $2.44 \times 10^{9}$ 
& $7.70 \times 10^{-2}$ 
\\[0mm]
(c)  
& $2.33 \times 10^{-64}$
& $4.62 \times 10^{-57}$
& $1.0 \times 10^{8}$
& $1.0 \times 10^{7}$ 
& $1.62 \times 10^{1}$ 
\\[0mm]
(d)  
& $2.33 \times 10^{-64}$
& $4.62 \times 10^{-57}$
& $1.0 \times 10^{6}$ 
& $1.0 \times 10^{5}$ 
& $1.62 \times 10^{1}$ 
\\[0mm] 
(e)  
& $2.85 \times 10^{-64}$
& $5.66 \times 10^{-57}$
& $1.0 \times 10^{4}$ 
& $8.0 \times 10^{2}$  
& $2.23 \times 10^{1}$ 
\\[0mm]
(f)  
& $2.85 \times 10^{-64}$
& $5.66 \times 10^{-57}$
& $1.0 \times 10^{2}$ 
& $8.0$  
& $2.23 \times 10^{1}$ 
\\
\botrule
\end{tabular}}
\label{table-1}
\end{table}

In Table~\ref{table-1}, 
we show the cases in which 
$B(H_0^{-1}, t_0) = {\cal O}(10^{-64})$ G 
on the Hubble horizon scale at the present time $t_0$ is generated 
for $X(t_k) = \exp \left( \chi_k \right)$ with $\chi_k = -0.940$, 
$g_{\mathrm{ps}} = 1.0$, 
$\xi_k = 2.5590616$, and 
$k = 2\pi/\left( 2997.9 h^{-1} \right) \, {\mathrm{Mpc}}^{-1}$ with $h=0.673$ 
(which is consistent with the recent Planck result~\cite{Planck:2015xua}). 
In the cases (j) (j = a, b, c, d, e, f), we obtain 
$C_+(k, t_{\mathrm{R}}) = (0.528, 0.528, 0.172, 0.172, 0.211, 0.211,)$, 
$T_\mathrm{R} \, [\mathrm{GeV}] = (1.02 \times 10^{14},\, 3.22 \times 10^{13},\, 3.22 \times 10^{12},\, 3.22 \times 10^{11},\, 3.22 \times 10^{10},\, 3.22 \times 10^{9})$, and 
$\bar{V}/M_\mathrm{P}^4 = (5.07 \times 10^{-15},\, 5.07 \times 10^{-17},\, 5.07 \times 10^{-21},\, 5.07 \times 10^{-25},\, 5.07 \times 10^{-29},\, 5.07 \times 10^{-33})$. 
It is confirmed that 
$C_+(k, t_{\mathrm{R}})$ is ${\cal O}(0.1)$ for the wide range of $H_{\mathrm{inf}}$ and $m$. 
All the results in Table~\ref{table-1} satisfy 
the constraints on the non-Gaussianity~\cite{Ade:2015ava} 
and the tensor-to-scalar ratio~\cite{Ade:2015lrj} from the Planck analysis. 
The magnetic field strength of ${\cal O}(10^{-64})$ G is also 
compatible with both the theoretical backreaction problem~\cite{Demozzi:2009fu} and the recent observational constraints from the Planck satellite.~\cite{Ade:2015cva}

\section{Non-Gaussianity of the curvature perturbations}

In the context of string theories, the gauge symmetry 
is spontaneously broken, so that the gauge field can have the mass. 
Therefore, we study such a spontaneous symmetry breaking 
by considering the case that the $\mathrm{U}(1)_Y$ gauge field 
has its coupling to other Higgs-like field $\varphi$, 
which can evolve to the vacuum expectation value. 
The kinetic term of $\varphi$ reads $\left| D \varphi 
\right|^2$, where the covariant derivative for $\varphi$ is defined by 
$D_\mu \equiv \partial_\mu + i g' F_\mu$ with $g'$ the gauge coupling~\cite{Meerburg:2012id}. 
The gauge field acquires the mass thanks to 
the Higgs mechanism in terms of $\varphi$. 
The quantum fluctuations of $\varphi$ lead to the quantum fluctuations on the mass of the gauge field. 
Hence, the amount of the quantum fluctuations 
is equal to that of quanta of the generated gauge field. 
Consequently, the gauge field generation yields 
the perturbations of number of $e$-folds during inflation $\delta N$. 
The local type non-Gaussinanity of 
anisotropy of the CMB radiation originates from these 
perturbations. 
The non-Gaussianity can be analyzed 
by exploring the curvature perturbations coming from 
the quantum fluctuations of $\varphi$. 
For the model of inflation in Ref.~\citenum{Linde:2012bt}, 
with the COBE normalization of power spectrum of the curvature perturbations~\cite{Smoot:1992td} 
$\Delta_{\mathcal{R}}^2 (k) = 2.4 \times 10^{-9}$ 
at $k=k_* = 0.002 \, \mathrm{Mpc}^{-1}$, 
the local type non-Gaussianity $f_\mathrm{NL}^\mathrm{local}$ 
is given by~\cite{Meerburg:2012id} 
$f_\mathrm{NL}^\mathrm{local} \approx 1.0 \times 10^{14} 
\left(g'{}^4 / \xi^{6}\right) \left(m^2/H_{\mathrm{inf}}^2\right)$.  

\begin{table}[t]
\tbl{Local type non-Gaussianity.}
{\begin{tabular}{@{}cccccc@{}}
\toprule 
& 
$f_\mathrm{NL}^\mathrm{local}$
& $g'{}^2$
& $H_{\mathrm{inf}} \hspace{1mm} [\mathrm{GeV}]$
& $m \hspace{1mm} [\mathrm{GeV}]$
& $Y_k/M_\mathrm{P}$ 
\\
\colrule
(A) 　
& 
$2.70$　
& $1.13 \times 10^{-5}$ 
& $1.0 \times 10^{13}$
& $2.44 \times 10^{12}$　
& $7.70 \times 10^{-2}$
\\[0mm] 
(B) 　
& 
$2.12 \times 10^{8}$　
& $1.0 \times 10^{-1}$ 
& $1.0 \times 10^{12}$
& $2.44 \times 10^{11}$　
& $7.70 \times 10^{-2}$
\\\botrule
\end{tabular}}
\label{table-2}
\end{table} 

In Table~\ref{table-2}, we list 
the local non-Gaussianity $f_\mathrm{NL}^\mathrm{local}$ of 
the curvature perturbations 
for $M =1.0 \times 10^{-1} M_\mathrm{P} = 2.43 \times 10^{17} \, 
\mathrm{GeV}$, $g_{\mathrm{ps}} = 1.0$, and 
$k = 2\pi/\left( 2997.9 h^{-1} \right) \, {\mathrm{Mpc}}^{-1}$ with $h=0.673$. 
Through the relation $\bar{V} = 3 H_{\mathrm{inf}}^2 M_\mathrm{P}^2$, 
we determine the value of $\bar{V}$ as 
$\bar{V} = 5.07 \times 10^{-11} M_\mathrm{P}^4$ for the case (A) 
and 
$\bar{V} = 5.07 \times 10^{-13} M_\mathrm{P}^4$ for the case (B). 
The value of $C_+(k, t_{\mathrm{R}})$ is 
(the case (A), the case (B)) = $(0.528, 0.528)$. 
The magnetic field strength on the 
Hubble horizon scale at the present time 
$B(H_0^{-1}, t_0) \hspace{1mm} [\mathrm{G}]$ is estimated as 
 (the case (A), the case (B)) = 
$(7.15 \times 10^{-64}, 7.15 \times 10^{-64})$, 
where we have used Eq.~(\ref{eq:3.35}) with the absolute value of $C_+(k, t_{\mathrm{R}})$. 
The constraint on $f_\mathrm{NL}^\mathrm{local}$ 
acquired from the Planck satellite~\cite{Ade:2015ava} 
is $f_\mathrm{NL}^\mathrm{local} = 2.7 \pm 5.8 \, (68 \% \, \mathrm{CL})$. 
It follows from Table~\ref{table-2} that the value of 
$f_\mathrm{NL}^\mathrm{local}$ in the case (A) can satisfy the constraints of 
the Planck results, but that of $f_\mathrm{NL}^\mathrm{local}$ 
in the case (B) cannot. 
The parameter sets whose values are close to those of the case (A) 
in Table~\ref{table-2} can also meet 
the constraints on $f_\mathrm{NL}^\mathrm{local}$, 
although the parameter space to meet 
the constraints of $f_\mathrm{NL}^\mathrm{local}$ is small.

\section{Tensor-to-scalar ratio of the density perturbations}

The definition of the tensor-to-scalar ratio $r$ is 
the amplitude of tensor modes (i.e., the primordial gravitational waves) 
of the density perturbations divided by 
that of their scalar modes. 
With a slow-roll parameter 
$\epsilon \equiv \left(M_\mathrm{P}^2/2\right) 
\left[ \left(dV (Y)/dY\right)/V(Y) \right]^2$, 
$r$ is represented as~\cite{Barnaby:2011vw} 
$r= 16 \epsilon (t_k)$, where $\epsilon (t_k) = \epsilon (t=t_k) = 
2M_\mathrm{P}^2 m^4 Y_k^2/\left(2\bar{V} - m^2 Y_k \right)^2$.  
In Table~\ref{table-3}, 
we display the values of the tensor-to-scalar ratio $r$ 
in the cases (A) and (B) (which are the same in Table~\ref{table-2}). 
The upper bound estimated by the Planck analysis is 
$r < 0.11 (95 \% \, \mathrm{CL})$.~\cite{Ade:2015lrj} 
It is clearly seen that this upper limit can be met in 
these cases. 

Thus, if the magnetic fields on the Hubble horizon scale with 
the current field strength ${\cal O}(10^{-64})$ G are generated, 
not only the local non-Gaussianity but also the tensor-to-scalar ratio 
in terms of the CMB radiation, which are compatible with the recent 
Planck data, can be produced for a certain parameter space. 

\begin{table}[t]
\tbl{Tensor-to-scalar ratio.}
{\begin{tabular}{@{}ccccc@{}}
\toprule
& $r$
& $H_{\mathrm{inf}} \hspace{1mm} [\mathrm{GeV}]$
& $m \hspace{1mm} [\mathrm{GeV}]$
& $Y_k/M_\mathrm{P}$ 
\\\colrule
(A)  
& $1.87 \times 10^{-5}$
& $1.0 \times 10^{13}$
& $2.44 \times 10^{12}$
& $7.70 \times 10^{-2}$
\\[0mm]
(B)  
& $1.87 \times 10^{-5}$ 
& $1.0 \times 10^{12}$
& $2.44 \times 10^{11}$
& $7.70 \times 10^{-2}$
\\
\botrule
\end{tabular}}
\label{table-3}
\end{table}

\section{Summary} 

In the present paper, we have investigated the generation of the large-scale magnetic fields from a kind of moduli inflation. 
We have first analyzed the values of three cosmological observables: Current strength of the magnetic fields on the Hubble horizon scale, local non-Gaussianity, and the tensor-to-scalar ratio. 
We have found that the local non-Gaussianity and tensor-to-scalar ratio obtained in this model can satisfy the recent constraints acquired from the Planck satellite.

\section*{Acknowledgments}

The author would like to sincerely appreciate the significant discussions with Professor Tatsuo Kobayashi and Professor Osamu Seto and their kind meaningful suggestions and comments. Furthermore, he is really grateful to 
Professor Akio Sugamoto and Professor Masahiro Morikawa for their very warm 
advice and discussions. This work was partially supported by the JSPS Grant-in-Aid for Young Scientists (B) \# 25800136 and the research-funds supplied from Fukushima University.




\end{document}